# Tikuna: An Ethereum Blockchain Network Security Monitoring System

Andres Gomez Ramirez[1,2] [✉], Loui Al Sardy[2,3] [✉], and Francis Gomez Ramirez[1,2]

[1] Edenia, Edificio Trifami, 10104 San José, Costa Rica
andres@edenia.com
[2] Sakundi, Sepapaja tn 6, 15551 Tallinn, Estonia
andres.gomez@sakundi.io
[3] Friedrich-Alexander-Universität Erlangen-Nürnberg, Faculty of Engineering,
Department of Computer Science, Martensstr. 3, 91058 Erlangen, Germany
loui.alsardy@fau.de

**Abstract.** Blockchain security is becoming increasingly relevant in today's cyberspace as it extends its influence in many industries. This paper focuses on protecting the lowest level layer in the blockchain, particularly the P2P network that allows the nodes to communicate and share information. The P2P network layer may be vulnerable to several families of attacks, such as Distributed Denial of Service (DDoS), eclipse attacks, or Sybil attacks. This layer is prone to threats inherited from traditional P2P networks, and it must be analyzed and understood by collecting data and extracting insights from the network behavior to reduce those risks. We introduce Tikuna, an open-source tool for monitoring and detecting potential attacks on the Ethereum blockchain P2P network at an early stage. Tikuna employs an unsupervised Long Short Term Memory (LSTM) method based on Recurrent Neural Network (RNNs) to detect attacks and alert users. Empirical results indicate that the proposed approach significantly improves detection performance, with the ability to detect and classify attacks, including eclipse attacks, Covert Flash attacks, and others that target the Ethereum blockchain P2P network layer, with high accuracy. Our research findings demonstrate that Tikuna is a valuable security tool for assisting operators to efficiently monitor and safeguard the status of Ethereum validators and the wider P2P network.

**Keywords:** Ethereum blockchain**,** security, P2P network, deep learning, anomaly detection, vulnerabilities, eclipse attacks.

## 1  Introduction

Ethereum was formally introduced by Vitalik Buterin in his whitepaper in 2014 [3] and launched in 2015 as a public cryptocurrency blockchain platform that supports smart contract functionality with Ether (ETH or Ξ) as its native cryptocurrency and Solidity as its programming language [36]; it is the second largest cryptocurrency after Bitcoin, with around $200 billion as of March 2023 [7, 38].

Even though blockchain technology is highly secure and decentralized, it still offers attack opportunities. For example, in blockchain networks, there are cases, such as the ones mentioned in [26, 8, 20], in which the dApps, average users, or the network itself are exposed to risks due to particular vulnerabilities [4, 40, 21, 37, 8, 39, 24, 26, 20]. Therefore, understanding the risks associated with blockchain networks and effectively developing security-focused solutions is essential to any blockchain.

Peer-to-peer (P2P) networks are decentralized networks that include many nodes storing and distributing data collectively, and each node operates as an individual peer.

The communication is carried out without a central authority; hence, all nodes obtain the same amount of power and are responsible for the same activities. The P2P network is one of the fundamental components of the blockchains that enable the creation and operation of cryptocurrencies [28].

In the blockchain, the P2P network enables nodes (clients) to exchange data, for instance, transactions and blocks. In general, there is an economic incentive for participants to behave honestly. Given their public and distributed nature, blockchain components are especially exposed to attackers who can easily reach and interact with the different layers. Such adversaries may use a malicious node, tool, or software to take advantage of specific weaknesses in the P2P network layer and launch several attacks on the blockchain, like the ones described in [26, 40, 20]. The security of the entire blockchain relies on the reliability of its P2P network.

The Ethereum P2P protocol [35] was influenced by the kademlia Distributed Hash Table (DHT) design. Although kademlia possesses valuable properties, it has several limitations in terms of its security [4, 22]. There are several known attacks for such a protocol, including the eclipse attacks [20, 40], where it is possible to perform manipulations against the Ethernet P2P network participants, and deanonymization attacks, as presented in [14]. Other types of vulnerabilities are also present (s. Section 3.2). Nevertheless, employing multiple detection and mitigation approaches [10, 11] can significantly reduce or eliminate the severity of these risks. For example, monitoring the health of the blockchain network by collecting data and deriving insights could assist in the detection process of different incidents as well as enhance network status visibility. In addition, it could help devise future mitigation approaches for these kinds of attacks.

This research paper introduces the following three main contributions:
- A Machine Learning (ML) approach that can detect several attacks at the Ethereum P2P layer using peer message trace data in a testing simulation environment using the libp2p testground framework;
- The detection of eclipse attacks on the mainnet is demonstrated by extracting custom-generated discovery connection log data from the Ethereum client Prysm and utilizing LSTM neural network;
- A custom exploit of an eclipse attack was developed and tested against a modified Prysm client on the mainnet. The peer table buckets could be fulfilled by a single attacking machine, overcoming the limitation of a single peer per IP address by using virtual addresses and Docker containers. With this exploit, the effectiveness of the Tikuna approach can be tested.

This paper is organized as follows: Section 2 provides an overview of alternative and related approaches. Next, in Section 3, the various types of blockchain P2P network attacks are discussed, and the Tikuna approach, consisting of three primary steps, is introduced. The efficacy of the Tikuna approach is evaluated in Section 4, utilizing a simulation and mainnet connection dataset. Finally, Section 5 concludes the paper by summarizing the proposed work, drawing conclusions, and identifying potential future research directions.

## 2   Related Work

Peer-to-peer (P2P) networking is the component that serves as the foundation for any blockchain technology. But, similar to other forms of technology, P2P networks may be affected by security issues. Adversaries can exploit some of these vulnerabilities to carry out several attack vectors on the blockchain, such as eclipse attacks, selfish mining attacks, Sybil attacks, or DDoS attacks. Therefore, researchers have recently started focusing on this area to address the different attack vectors on the Ethereum platform and the P2P network security vulnerabilities.

The following are some of the most recent works that address the security challenges of the Ethereum blockchain P2P networks:

Kabla et al. [21] focus on the security issues of each layer in the Ethereum blockchain, such as the network layer, by providing an in-depth analysis covering the following three areas:
- Its potential attacks include eclipse attacks and account hijacking attacks.
- The vulnerabilities that lead to them are unlimited node creation and uncapped incoming connections.
- Each incident's consequences include double spending or Denial of Service (DoS).

Furthermore, the work presents an overview of the effectiveness and limitations of the current Intrusion Detection Systems (IDS) as a defense technique against various Ethereum-based attacks.

Vyzovitis et al. [37] propose two different hardening measures for the GossipSub protocol, the mesh construction and the score function. GossipSub is a messaging protocol that enables rapid and robust message transmission in permissionless networks, such as the Ethereum 2.0 network layer, to make it, on the one hand, more secure against various types of attacks (such as the sybil attack, the eclipse attack, and the Cold Boot attack), and on the other hand, to facilitate a faster transmission of the messages within the network. In addition, the authors describe some of the countermeasures featured in the GossipSub protocol, such as controlled mesh maintenance, opportunistic grafting, and adaptive gossip distribution. The writers evaluate the GossipSub protocol against various attacks and demonstrate its resistance to these attacks. However, the proposed methods use fixed rules that should be manually parametrized, which has limited their widespread usage in the different Ethereum clients. We suggest the use of machine learning to automatically select parameters for the detection of attacks.

The report from Least Authority [23] details the results of a security audit they conducted on the next-generation node discovery protocol of the Ethereum P2P network stack. Their evaluation focused on network operations, potential attacks, and the protocol's cryptography. The report reveals areas for improvement in the DevP2P specification, particularly the lack of a proof-of-X scheme for identity generation, disjoint paths in the lookup operation, and broken handshake authentication. Furthermore, the status of each issue and suggestion is indicated, with some being resolved and others unresolved or in discussion. Finally, the report strongly recommends implementing an identity-proof system in the next version of the Ethereum node discovery protocol. The report also indicates that launching eclipse attacks against the Ethereum clients using the current peer discovery specification is trivial.

Marcus et al. [24] highlight the possibility of eclipse attacks on Ethereum nodes, which could be carried out using only two hosts and could result in the victim's view of the blockchain being filtered or their computing power being co-opted. The authors' contributions include a detailed explanation of the network and its relationship with the kademlia protocol, two off-path eclipse attacks, and one involving time manipulation. Furthermore, they have proposed countermeasures to prevent these attacks, such as using a combination of IP address and public key for node identification and making design decisions to harden Ethereum. Some of these countermeasures have been implemented in Geth v1.8. Those measures restrict the number of peers connecting to a victim from the same IP. We show that it is still possible to fulfill buckets from the peer table from a single attacking server with a unique public IP address.

Xu et al. [40] discuss the eclipse attacks on the Ethereum P2P Network. An eclipse attack is an attack that allows an adversary to isolate a target node within the P2P network by gaining complete control of a node's access to information or control over everything that the node sees. The authors developed an ETH-EDS eclipse-attack detection model targeting the Ethereum platform. This model used a random forest classification technique to examine the network's regular and attack data packets. The collected data packets included details like the size of the tag packets, the frequency with which they were accessed, and the access time. The findings of the experiments show that malicious network nodes could be identified with a high degree of precision. We further propose using deep learning techniques to automatically select features in the data and improve detection accuracy. We use this research to compare our results. The details of our approach are discussed in the following sections.

## 3   Tikuna Approach

### 3.1   Tikuna Terminology

Tikuna is a proof-of-concept peer-to-peer network security monitoring system developed initially for the Ethereum blockchain. It uses deep learning to extract security and performance insights for the early detection of incidents. Our goal with Tikuna is to support the Ethereum community by providing a cutting-edge open-source tool capable of collecting security-related data from the state of the P2P network and improving network visibility by providing insights about the network's current state.

The Ethereum peer-to-peer (P2P) discovery protocol [35] enables nodes on the network to locate and connect with other peers. With this protocol, nodes on the Ethereum network can share information about transactions, blocks, and other network events. The DevP2P architecture includes the discovery protocol as an essential component of the communication system among Ethereum nodes.

Ethereum uses a discovery algorithm similar to Kademlia [25], a Distributed Hash Table (DHT) communication protocol used before for other technologies such as torrents. This protocol enables peers to identify and interact with each other in a decentralized network without having to rely on a central server. Every node in the network is responsible for its own routing table, organized in the form of a binary tree with the node's ID at the tree's root. Other peers are listed as leaf nodes. An existing peer can assist a new peer in joining the network by checking its routing table to locate the node relatively closest to the new peer's ID. This is accomplished by utilizing a distance metric based on the peer IDs' XOR operation. This process of gathering information about other peers in the network is repeated iteratively until the new peer has collected data on a significant number of peers in the network. The distance metric is the reason for both the effectiveness and the scalability of kademlia's routing tables, even when applied to extremely large networks.

The unsupervised anomaly detection method selected for this work is the long short term memory. These algorithms are commonly used for analyzing time series data and natural language processing. Below is a brief introduction to these neural network algorithms.

**Recurrent Neural Network.** Recurrent neural network [26] is a type of neural network frequently utilized for processing sequential data, such as time series. RNN is specialized for processing a sequence of values that are a function of time. We can define a data sequence as follows:

$$x(1), \ldots, x(T) \qquad (1)$$

, where $T$ is the number of available data samples. RNN can scale to long sequences that would not be practical for networks without sequence-based specialization. Most recurrent networks can also process sequences of variable length. One of these models is especially interesting for this research. The long short term memory model [5, 29, 33, 26, 1] uses a gating mechanism to propagate information through many time steps properly. LSTM networks have a specific memory cell and can capture long-term dependencies in sequential data. They are valuable tools for language modeling problems. These models are a version of recurrent neural networks useful for long interrelated sequences of data [5, 29, 33, 26, 1]. LSTM was chosen in this research for anomaly detection to find malicious connections to an Ethereum client. They can be defined with the following set of equations:

$$\vec{f}_t = \sigma_g(W_f \vec{x}_t + U_f \vec{h}_{t-1} + \vec{b}_f) \tag{2}$$

$$\vec{\imath}_t = \sigma_g(W_i \vec{x}_t + U_i \vec{h}_{t-1} + \vec{b}_i) \tag{3}$$

$$\vec{o}_t = \sigma_g(W_o \vec{x}_t + U_o \vec{h}_{t-1} + \vec{b}_o) \tag{4}$$

$$\vec{c}_t = \vec{f}_t \circ \vec{c}_{t-1} + \vec{\imath}_t \circ \sigma_c(W_c \vec{x}_t + U_c \vec{h}_{t-1} + \vec{b}_c) \tag{5}$$

$$\vec{h}_t = o_t \circ \sigma_h(c_t) \tag{6}$$

Similarly to the common RNN, $\vec{x}_t$ is the input vector at a given iteration $t$, $\vec{h}_t$ is an output vector of the hidden layer, and $\vec{c}_t$ is a cell state. In this case, $W$ and $U$ are parameter matrices, and $\vec{b}$ are bias vectors. $\vec{f}_t$ is a forget gate vector, $\vec{\imath}_t$ is the input gate vector and $\vec{o}_t$ is the output gate vector. The operator $\circ$ is the entrywise product of matrices.

In the next section, some of the attacks that can be detected using the described unsupervised anomaly detection model are explained in detail.

### 3.2 Types of P2P network attacks

Adversaries can exploit some vulnerabilities in the blockchain's P2P networks to perform a variety of attacks [4, 40, 21, 37, 8, 39, 24, 26, 20], including the following:

(1) **Eclipse Attack [39, 24, 20].** An eclipse attack is an attack that can be carried out against a single victim node or the whole network, where the adversary isolates the victim node within the P2P network by gaining complete control of the node's access to information or control over everything that the node sees.
(2) **Censorship Attack [37].** During this type of attack, the adversaries will use the nodes on the network that they have created with fake identities (i.e., Sybil nodes) to propagate all messages, except for those that the peer published that they are trying to attack. In addition, the primary objective of the attacker is to censor the target and stop its messages from being transmitted to the rest of the network.
(3) **Sybil Attack [12, 2].** Which is also known as pseudo-spoofing, is an attack that can target any P2P network, such as blockchain networks, in which a single adversary creates a large number of nodes on the network with fake identities to gain a more significant presence in the network and eventually take control of the network. This kind of attack might also be used to carry out other types of attacks, such as an eclipse or censorship attack.
(4) **Cold Boot Attack [37].** In this type of attack, honest nodes and nodes with fake identities (so-called Sybil nodes) join the network simultaneously; genuine peers attempt to build their network while connecting to both Sybil and genuine peers. Since there is no information about honest nodes to secure the network, the Sybils can seize control. There are two possible scenarios for the attack: (1) when the network bootstraps with Sybils joining from the start, or (2) when new nodes join the network when it is under attack.

(5) **Flash and Covert Flash Attack [37].** Sybils will simultaneously connect and launch attacks against the targeted network in a Flash attack. On the other hand, in the Covert Flash Attack, Sybils join the network and act normally for some time to build up their score. Then, they carry out a coordinated attack in which they stop propagating messages altogether to disrupt the network entirely. Furthermore, as the adversaries act appropriately up until that point and establish a valid profile, it is difficult to identify the attack.

Our goal with Tikuna is to identify the described attacks using the anomaly detection approach, that is, by finding peer connections to a victim that deviate from the expected behavior of honest peers. We describe in detail the different components of Tikuna, starting with the data collection and finalizing with the anomaly detection module.

### 3.3 Tikuna Methodology

Fig. 1 shows the methodology of Tikuna, which comprises three main steps: (1) data extraction from a simulation environment using the testground [34] framework and the Ethereum mainnet; (2) training and classification analysis; and (3) P2P security incident detection. The following subsections provide a detailed explanation of these steps.

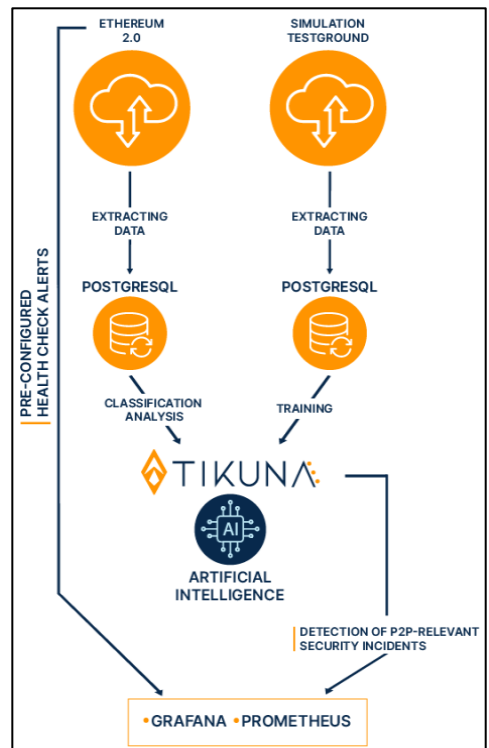

**Fig. 1.** Overview of the proposed Tikuna Architecture.

**Step 1: Data extraction from testground simulation and Ethereum 2.0 mainnet.**
Data extraction refers to extracting data from a simulation or mainnet environment. It may include patterns that are challenging to identify without suitable analysis and converting it into a format ideal for the training part, i.e., for training our LSTM model. However, before this step, the dataset must be preprocessed to extract the pertinent features and convert the data into a format the AI model can interpret. Every second, the measurement system gathers a sequence of monitoring data from the participating peers in the network. The extracted data is parsed into structured data represented by vectors of integers that are later normalized. This data includes timestamps, and gossip message event traces. The data extraction process in LSTM [5, 29, 33, 26, 1] involves four main steps: (1) the data cleaning step filters out any data from the simulation dataset considered irrelevant or corrupt, (2) the feature extraction step involves identifying and extracting the relevant features from the dataset, which will be used to train the LSTM model, (3) The data normalization step scales the extracted features to a standard range, ensuring that the LSTM model can handle them most effectively, and (4) in the sequence formation step, the extracted and normalized features are grouped into a time-series sequence that can be utilized to train the LSTM model.

**Step 2: Training and classification analysis.** The model is fed with input sequences and output labels corresponding to only normal data in the training phase. The model's weights and biases are then iteratively updated to reduce the difference between its predictions and the actual outputs. This enables the model to understand the underlying relationships in the data. In the evaluation part, on the other hand, the trained LSTM model is utilized to predict new, unseen input sequences. The model receives a sequence of input data and generates an output prediction based on the learning patterns during the training process. This prediction may then be compared to the actual label to determine the model's accuracy. As illustrated in Fig. 2, the training data for Tikuna AI are the output data from the pre-processing stage for regular peer communication within the network. In addition, Tikuna uses this data to train the model and extract features that the artificial neural network in the subsequent stage will utilize.

**Step 3: Detection of P2P-relevant security incidents.** In this step, detecting security incidents related to the P2P network involves identifying and recognizing connection patterns that characterize the threats described in Section 3.2. The goal is to quickly identify and respond to such incidents, minimize damage, and maintain network infrastructure security. As shown in Fig. 2, an LSTM method [5, 29, 33, 26, 1] is used by Tikuna. Such a model is based on a recurrent neural network, and it can remember long-term dependencies over the input data (i.e., a series of connection monitoring data). In addition, a forecasting loss function is used to evaluate how well the neural network models the training data by comparing the target and predicting output values to minimize this function (i.e., to train the model to detect anomalies based on previous observations under the assumption that honest peers monitoring data follow a consistent pattern). Consequently, Tikuna detects P2P-relevant security incidents when the peers' connection data deviates from typical behavior.

Finally, Fig. 2 displays the essential steps of the process flow of Tikuna that ensure the model is thoroughly trained and capable of precisely detecting data anomalies.

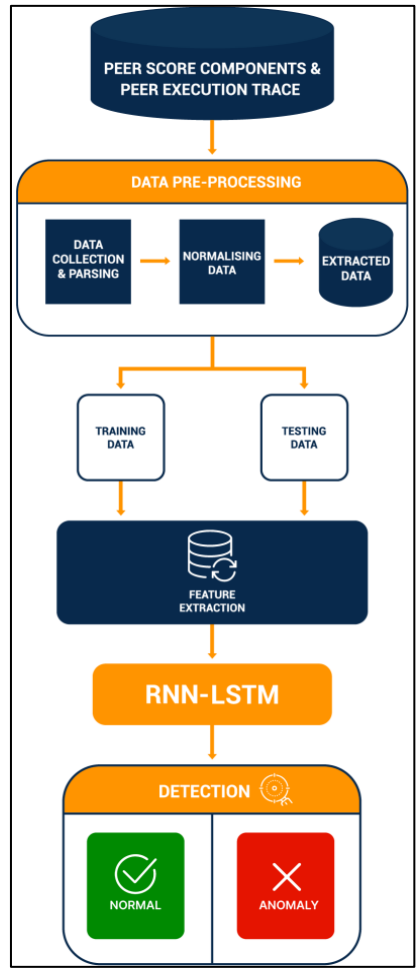

**Fig. 2.** Tikuna AI Flow Diagram.

## 4 Evaluation

### 4.1 Experiment Design

Experiments were conducted in two distinct network environments: one using the Protocol Labs simulation tool testground [34] and the other using the Ethereum mainnet to thoroughly evaluate the effectiveness and performance of the Tikuna approach. In this research, we utilized simulations to demonstrate that Tikuna is a practical approach to detecting Ethereum blockchain P2P network attacks.

For all the experiments performed, we have used a set of five dedicated root Hetzner servers in different locations worldwide. They all had 64 GB of DDR4 RAM, two 512GB NVMe SSDs, and an AMD Ryzen CPU as hardware characteristics. As mentioned before, we have only used the Ethereum mainnet client Prysm [31] because it is the most popular node software at the time of writing. In future research, we plan to explore other prevalent clients.

### 4.2 Attack Simulation Setup

Since we have two different environments for testing and the mainnet, we have used various strategies to simulate the attacks we wanted to detect. In the testing environment, we used this repository [18], created from a research project by Protocol Labs aimed to recreate several attacks on the libp2p (go-lang) library version, which is the one used by Ethereum (Prysm), Filecoin, and IPFS. All the attacks described in Section 3.2 are executed in such a simulation environment. We have forked the gossipsub-hardening repository [19] and modified it to store the peer message traces in a file. We also had to make some changes to be able to run the software since it was written a couple of years before this paper. A considerable amount of traces are produced during the simulation of the attacks; hence, we have grouped the traces between the 12 types [17] of gossip-sub events and the number of events seen every 300 milliseconds. In Fig. 3, we show samples of the kind of data used.

For the mainnet scenario, we have developed our own eclipse attack code to test the effectiveness of our detection approach in a more realistic environment. The exploit is based on the work in [9] using the Rust programming language. It uses the testground framework to run a series of Ethereum nodes that create fake node IDs specially crafted to be located in specific buckets of a victim's Ethereum client peer table.

With the developed exploit code, we could simulate a realistic eclipse attack scenario against a modified Prysm client (using the Geth discovery library). We have changed the code of both projects so the victim node will not advertise the simulated fake peer IDs to other honest peers in the network. We also added new logging features to collect the UDP discovery connections and the gossip-sub message traces received and sent by the victim client. The code forks are in the following repositories [32, 15]. Fig. 4 shows a sample of the collected discovery connection data from the honest and attacking peers. The data was collected from a single victim Ethereum node. Each line has several input features, including a timestamp, IP, and port removed from the peer table, IP and port added to the peer table, and bucket where the peer is added.

### 4.3 Deep Learning Algorithm Setup

With the developed exploit code, we could simulate a realistic eclipse attack scenario against a modified Prysm client (using the Geth discovery library). We have changed the code of both projects so the victim node will not advertise the simulated fake peer IDs to other honest peers in the network. We also added new logging features to collect the UDP discovery connections and the gossip-sub message traces received and sent by the victim client. The code forks are in the following repositories [32, 15]. Fig. 4 shows a sample of the collected UDP discovery connection data from the honest and attacking peers. The data was collected from a single victim Ethereum node. Each line has several input features, including a timestamp, IP, and port removed from the peer table, IP and port added to the peer table, and bucket where the peer is added.

| | timestamp | 0 | 2 | 3 | 4 | 5 | 6 | 7 | 8 | 9 | 11 | 12 | peer | honest |
|---|---|---|---|---|---|---|---|---|---|---|---|---|---|---|
| 22276 | 2022-11-10 21:48:04.000 | 0 | 0 | 0 | 0 | 0 | 0 | 0 | 0 | 1 | 0 | 0 | 1 | True |
| 24107 | 2022-11-10 21:48:04.300 | 0 | 0 | 0 | 2 | 0 | 0 | 0 | 0 | 0 | 0 | 0 | 1 | True |
| 22277 | 2022-11-10 21:48:04.400 | 0 | 0 | 0 | 1 | 0 | 3 | 0 | 0 | 0 | 0 | 0 | 1 | True |
| 22278 | 2022-11-10 21:48:04.500 | 0 | 0 | 0 | 7 | 0 | 6 | 0 | 0 | 0 | 0 | 0 | 1 | True |
| 22279 | 2022-11-10 21:48:04.600 | 0 | 0 | 0 | 1 | 0 | 4 | 0 | 0 | 0 | 2 | 0 | 1 | True |
| ... | ... | ... | ... | ... | ... | ... | ... | ... | ... | ... | ... | ... | ... | ... |
| 80605 | 2022-11-10 21:51:37.200 | 0 | 0 | 0 | 0 | 0 | 1 | 0 | 0 | 0 | 0 | 0 | 50 | True |
| 80606 | 2022-11-10 21:51:37.400 | 0 | 0 | 0 | 0 | 0 | 1 | 0 | 0 | 0 | 0 | 0 | 50 | True |
| 80537 | 2022-11-10 21:51:37.500 | 0 | 0 | 0 | 0 | 0 | 1 | 0 | 0 | 0 | 0 | 0 | 50 | True |
| 80633 | 2022-11-10 21:51:37.700 | 0 | 0 | 0 | 0 | 0 | 1 | 0 | 0 | 0 | 0 | 0 | 50 | True |

| | timestamp | 0 | 2 | 3 | 4 | 5 | 6 | 7 | 9 | 11 | peer | honest |
|---|---|---|---|---|---|---|---|---|---|---|---|---|
| 3652 | 2022-11-24 23:35:40.100 | 0 | 0 | 0 | 0 | 0 | 0 | 0 | 1 | 0 | 1 | False |
| 3653 | 2022-11-24 23:35:40.700 | 0 | 0 | 0 | 1 | 0 | 0 | 0 | 0 | 0 | 1 | False |
| 3654 | 2022-11-24 23:35:40.800 | 0 | 0 | 0 | 1 | 0 | 2 | 0 | 0 | 0 | 1 | False |
| 3655 | 2022-11-24 23:35:40.900 | 0 | 0 | 0 | 1 | 0 | 0 | 0 | 0 | 0 | 1 | False |
| 3656 | 2022-11-24 23:35:41.000 | 0 | 0 | 0 | 4 | 1 | 3 | 0 | 0 | 0 | 1 | False |
| ... | ... | ... | ... | ... | ... | ... | ... | ... | ... | ... | ... | ... |
| 5403 | 2022-11-24 23:39:04.400 | 1 | 141 | 7 | 0 | 0 | 151 | 81 | 0 | 0 | 1 | False |
| 5404 | 2022-11-24 23:39:04.500 | 0 | 98 | 19 | 0 | 0 | 117 | 234 | 0 | 0 | 1 | False |
| 5405 | 2022-11-24 23:39:04.600 | 1 | 54 | 5 | 0 | 0 | 58 | 56 | 0 | 0 | 1 | False |
| 5406 | 2022-11-24 23:39:04.700 | 0 | 150 | 15 | 0 | 0 | 166 | 166 | 0 | 0 | 1 | False |
| 5407 | 2022-11-24 23:39:04.800 | 1 | 2 | 9 | 0 | 0 | 11 | 101 | 0 | 0 | 1 | False |

**Fig. 3.** Example data extracted from testground simulations.

Forecasting loss was utilized to model the sequences of peer traces and connection log data and predict the subsequent observed event using the previous observations. By learning event patterns from regular series, we could automatically detect anomalies when the event pattern deviates from the ordinary operation [5]. We divide the data into fixed-length sequences to give the machine learning algorithm its inputs. Each input sequence should correspond to a single output label, in our case, the following token in the sequence. Then we needed to transform input sequences into tensors.

The tensors should have the shape (batch_size, time_steps, input_features), where batch_size represents the number of input sequences in a single batch, time_steps represents the length of each input sequence, and input_features represents the number of features in each input data point.

|  | Timestamp | Removed IP | Removed Port | Added IP | Added Port | Bucket | label |
| --- | --- | --- | --- | --- | --- | --- | --- |
| 0 | [2023-01-28\|01:13:41.042] | 162.55.1.114 | 11900 | 83.151.202.176 | 1024 | 256 | normal |
| 1 | [2023-01-28\|01:13:41.042] | 3.80.226.93 | 9000 | 114.156.141.196 | 12000 | 256 | normal |
| 2 | [2023-01-28\|01:13:41.042] | 68.170.92.177 | 52594 | 139.144.21.173 | 9000 | 256 | normal |
| 3 | [2023-01-28\|01:13:41.042] | 54.209.226.233 | 9000 | 135.181.178.95 | 9000 | 256 | normal |
| 4 | [2023-01-28\|01:13:41.042] | 18.224.59.77 | 9000 | 34.234.215.244 | 9000 | 256 | normal |
| ... | ... | ... | ... | ... | ... | ... | ... |
| 3376889 | [2023-01-31\|20:14:44.961] | 167.235.13.35 | 4000 | 34.230.92.127 | 37906 | 256 | normal |
| 3376890 | [2023-01-31\|20:14:44.961] | 78.46.89.12 | 30303 | 65.108.2.241 | 9505 | 256 | normal |
| 3376891 | [2023-01-31\|20:14:44.961] | 131.153.182.102 | 12000 | 85.17.132.26 | 12000 | 256 | normal |
| 3376892 | [2023-01-31\|20:14:44.961] | 70.184.72.180 | 9000 | 18.216.91.167 | 12000 | 256 | normal |
| 3376893 | [2023-01-31\|20:14:44.961] | 34.146.226.24 | 15360 | 142.132.140.121 | 12001 | 256 | normal |

|  | Timestamp | Removed IP | Removed Port | Added IP | Added Port | Bucket | label |
| --- | --- | --- | --- | --- | --- | --- | --- |
| 0 | [2023-02-03\|21:53:11.906] | 149.56.240.35 | 9000 | 16.0.186.130 | 9000 | 252 | abnormal |
| 1 | [2023-02-03\|21:53:11.907] | 35.207.99.26 | 9000 | 16.0.53.120 | 9000 | 255 | abnormal |
| 2 | [2023-02-03\|21:53:11.909] | 43.135.40.73 | 12000 | 16.0.160.76 | 9000 | 252 | abnormal |
| 3 | [2023-02-03\|21:53:11.909] | 100.27.30.226 | 9000 | 16.0.65.129 | 9000 | 256 | abnormal |
| 4 | [2023-02-03\|21:53:11.910] | 34.229.79.57 | 39085 | 16.0.170.200 | 9000 | 255 | abnormal |
| ... | ... | ... | ... | ... | ... | ... | ... |
| 1420 | [2023-02-04\|05:07:00.092] | 16.0.61.29 | 12000 | 205.185.120.171 | 12651 | 254 | abnormal |
| 1421 | [2023-02-04\|07:51:06.105] | 54.238.108.184 | 33311 | 16.0.61.29 | 12000 | 254 | abnormal |
| 1422 | [2023-02-04\|07:51:38.907] | 16.0.61.29 | 12000 | 162.55.134.100 | 49429 | 254 | abnormal |

**Fig. 4.** Sample of normal and eclipse attack mainnet data

Formally, for an event $e_i$ at time step $t$, an input window W is created, which contains $m$ connection events preceding $e_i$, i.e., $W = [e_{t-m}, ..., e_{t-2}, e_{t-1}]$. This is achieved by splitting event sequences into subsequences. Window size and step size are the parameters that control the division process.

The model is then trained to learn a conditional probability distribution $P(e_t = e_i | W)$ for all $e_i$ in the set of distinct log events $E = \{e_1, e_2, ..., e_n\}$. In the detection phase, the trained model predicts a new input window, which will be compared against the actual event. An anomaly is seen if the ground truth is not one of the most $k$ probable events predicted by the model.

Given the numerical labels, the trace data collected in the testground simulation attacks required a mean squared error (squared $L2$ norm) loss function. On the other hand, the discovery connection data collected from the mainnet attacks required a cross-entropy loss function because of the categorical labels (the most probable following tokens).

Table 1. Parameters selected for the LSTM model. summarizes the various parameters that may be adjusted in the LSTM model for the specific type of data modeled. The hidden_size, num_layers, num_directions, and the embedding_dim were all fixed, and the suggested model defined the values for each parameter. The parameters max_token_len, min_token_count, epochs, batch_size, learning_rate, topk, patience, and random_seed had their values predetermined, and the relevant experimental experience was used to identify their appropriate ranges.

Table 1. Parameters selected for the LSTM model.

| Parameters/ Data type | Testground trace data | Mainnet discovery connection data |
|---|---|---|
| hidden_size | 20 | 128 |
| num_layers | 2 | 2 |
| num_directions | 2 | 2 |
| embedding_dim | 5 | 10 |
| epoches | 100 | 100 |
| batch_size | 1000 | 1024 |
| learning_rate | 0.01 | 0.01 |
| topk | - | 5 |
| patience | 5 | 30 |
| random_seed | 50 | 42 |

### 4.4 Experiment Results

Regarding the eclipse attack on a mainnet client, it was possible to overcome the Prysm restriction by adding many nodes from the same public IP address into the same peer table bucket. We used the ECDSA signatures using the secp256k1 curve to generate fake peer IDs and craft many Ethereum Node Records (ENR) for nodes that communicated with the victim's Prysm client. The exploit code will be published once it is reviewed by the Ethereum Foundation to confirm whether a fix is needed.

We include in this paper the ML detection results for three different attacks in the testground simulation environment: (1) multiple Sybil nodes launching eclipse attacks against a single node; (2) various nodes trying covert attacks against several honest peers; and (3) several attackers trying to eclipse an entire peer network. For the mainnet environment, we show the detection results for multiple nodes trying to eclipse a single victim node, and we compare the results with a previous approach using random forest classification over network packets [29]. Refer to Section 3.2 for an explanation of such attacks.

The results include standard measures like precision, recall, F1 score, and accuracy to evaluate the models with the different data types. The precision represents the proportion of true anomalies among all identified anomalies and can be calculated using the following equation:

$$precision = \frac{TP}{TP + FN} \quad (7)$$

, where TP stands for true positive. The recall metric determines the proportion of outliers detected in a data set (provided that the ground truth is known) and can be computed using the following equation:

$$Recall = \frac{TP}{TP + FN} \quad (8)$$

The F1 score is a metric that combines precision and recall to provide a single score that indicates how well the model is performing, and it is estimated using the following equation:

$$F1\ score = \frac{2 * precision * recall}{precision + recall} \quad (9)$$

Accuracy is the ability to correctly differentiate between anomalies and normal data and may be determined using the following equation:

$$Accuracy = \frac{TP + TN}{TP + TN + FP + FN} \quad (10)$$

Table 2. presents the results of applying our Tikuna anomaly detection approach for detecting attacks in simulated testground runs, including the described metrics, the number of attackers, and the number of victims.

Table 2. Summary of Tikuna results using the simulation test data

| Attack / Metric | Attackers | Victims | Precision | Recall | F1 score | Accuracy |
|---|---|---|---|---|---|---|
| **Eclipse Single Victim** | 100 | 1 | 1.00 | 0.99 | 0.99 | 0.99 |
| **Covert Attack** | 100 | 20 | 1.00 | 0.80 | 0.89 | 0.80 |
| **Eclipse Network** | 200 | 50 | 1.00 | 0.79 | 0.88 | 0.79 |

The results were collected after several LSTM iterations with training and evaluation data. As can be seen in Table 2. , the best results were obtained for the multiple attacker single victim scenario, with metrics close to 100% of performance. For the other two scenarios, the metrics indicate a lesser optimal performance, especially in recall and accuracy metrics, but still, our approach shows good detection ability.

Table 3. presents the results of applying our Tikuna approach to the Ethereum mainnet discovery connection data, including precision, recall, F1 score, and accuracy.

Table 3. Summary of Tikuna results using the Ethereum mainnet

| Approach / Metric | Precision | Recall | F1 score | Accuracy |
|---|---|---|---|---|
| **Tikuna** | 0.81 | 0.88 | 0.85 | 0.87 |
| **RFC** | 0.71 | 0.95 | 0.62 | - |

Four Hetzner servers were used for creating attacking Ethereum nodes, and one was used as a victim node. Except for recall, the Tikuna LSTM anomaly detection approach presented better results than the comparable work in [40], using Random Forest Classification (RFC) over network packets in all the metrics, especially the F1 metric that represents a better balance among true and false positives. The recall measure was the only metric where the RFC work performed better.

If we compare the results from the testground environment to the mainnet one, more optimal results were obtained for the simulation case with connection trace data. However, that same approach did not work for mainnet detection. Furthermore, the selected discovery connection log data model showed good performance, making it appropriate for usage in Ethereum blockchain validators.

## 5   Conclusion and Future Work

This paper presents Tikuna, an Ethereum blockchain network security monitoring and anomaly detection system, using a long short term memory-based neural network model. We introduced three main contributions: our method can detect several attacks at the P2P layer using peer message trace data in a testing simulation environment using the testground tool. We demonstrate the detection of eclipse attacks on the Ethereum mainnet by extracting discovery connection log data from the Prysm client. In addition, a custom exploit implementing an eclipse attack was developed and tested against a modified Prysm client on the mainnet.

Tikuna learns and encodes the expected behavior and the interaction between peers within the network, including timestamps, gossip-sub connection features, and discovery connection log data. It tries to classify this data as normal or malicious based on several attack patterns, such as eclipse and Covert attacks. Moreover, we presented the results of applying our approach to the Ethereum P2P network. We still need to work on reducing the number of false positives in the detection task, a classical problem faced by ML-based intrusion detection systems.

In future work, our team will continue with the development of Tikuna. Our ongoing efforts will be focused on identifying additional attacks, minimizing false positives, detecting real-world incidents, and incorporating different Ethereum clients. Moreover, we intend to investigate other areas of research where Tikuna can be utilized, such as Maximal Extractable Value (MEV). Finally, we will explore using our approach in other P2P networks based on the same technology and libraries used by Ethereum, like Filecoin and IPFS.

**Acknowledgment.** The authors gratefully acknowledge that the Ethereum Foundation Academic Research Grants supported the work presented. We also acknowledge all the support and helpful suggestions from our colleagues on the Edenia team.
.